\documentclass[10pt,twocolumn,letterpaper]{IEEEtran}

\usepackage{amsmath}
\usepackage{graphicx}
\include{psfig}
\usepackage{epsfig}
\usepackage{array}
\usepackage{psfrag}
\usepackage{amssymb}
\usepackage{color}
\usepackage{pstricks,pst-node,pst-text,pst-3d,pst-plot}
\usepackage{cite}

\usepackage{amsmath}
\usepackage{graphics}
\usepackage{trig}
\usepackage{graphicx}
\usepackage{amsmath, amsthm, amssymb}
\usepackage{anysize}
\usepackage{epsf}
\usepackage{psfrag}
\usepackage{pstricks}
\usepackage{pst-node}
\usepackage{pst-3d}
\usepackage{pst-plot}
\usepackage{epstopdf}

\usepackage{array}
\usepackage{amssymb}
\usepackage{color}

\usepackage{amsmath, amssymb, multirow, cite , epsfig, epstopdf, graphicx , graphics} %epsfig,
\usepackage{graphics}
\usepackage{verbatim}
\usepackage{subfigure}
\usepackage{algorithmicx}
\usepackage{algorithm}
\usepackage{algpseudocode}
\usepackage{algpascal}
\usepackage{algc, xcolor}

\ifCLASSINFOpdf
\else
\fi

\newcommand\AENDSKIP{\end{minipage}\bigskip}
\newcommand\AEND{\end{minipage}}

\begin{document}

\title{A Feature Selection Method Based on Shapley Value to False Alarm Reduction in ICUs, A Genetic-Algorithm Approach}
%\doublespacing

%\author{Mohammad Zaeri-Amirani,
%Fatemeh Afghah, Sherali Zeadally$^*$\thanks {\small F.~Afghah is the corresponding
%author. M.~Zaeri-Amirani and Fatemeh Afghah are with the School of Informatics, Computing and Cyber Systems, Northern Arizona University, PO Box 5698, Flagstaff, AZ 86011,
%USA, email: \texttt{\{Mohammad.Zaeri-Amirani, fatemeh.afghah\}@nau.edu}, phone: +1
%928-523-5095, S.~Zeadally is with the
%College of Communication and Information at the University of Kentucky, Lexington, KY 40506.
%USA, e-mail: \texttt{szeadally@uky.edu}, phone: +1 859-218-2299.}}

\author{
\IEEEauthorblockN{Mohammad Zaeri-Amirani, Fatemeh Afghah, Sajad Mousavi
\\}
\IEEEauthorblockA{School of Informatics, Computing and Cyber Systems,\\ Northern Arizona University,
Flagstaff, AZ 86011\\
Email:\{mohammad.zaeri-amirani,fatemeh.afghah, sajadmousavi
\}@nau.edu} %<------ Line breaks in the current column
%<------- Extra vertical space
}
\maketitle
\begin{abstract} High false alarm rate in intensive care units
(ICUs) has been identified as one of the most critical
medical challenges in recent years. This often results
in overwhelming the clinical staff by numerous false or
unurgent alarms and decreasing the quality of care through
enhancing the probability of missing true alarms as well as
causing delirium, stress, sleep deprivation and depressed
immune systems for patients. One major cause of false
alarms in clinical practice is that the collected signals from
different devices are processed individually to trigger an
alarm, while there exists a considerable chance that the
signal collected from one device is corrupted by noise or motion
artifacts. In this paper, we propose a low-computational
complexity yet accurate game-theoretic feature selection method which is based on a genetic algorithm that identifies the most informative biomarkers across the signals collected from various monitoring devices and can considerably reduce the rate of false alarms \footnote{This work was partially sponsored by the National Science Foundation under grant 1657260. }.
\end{abstract}
%Index Terms- Coalition games, feature selection, genetic algorithm, Shapley value,  sensitivity versus specificity level

\section{Introduction}
False alarms are widely considered the number one hazard imposed by the use of medical technologies. The Emergency Care Research Institute (ECRI) named alarm hazards as number 1 of the "Top 10 Health Technology Hazards" for several years \cite{ECRI15}. These false alarms can be due to several factors such as low threshold setting of the monitoring devices, motion artifacts, and sensor detachment or malfunction causing \emph{alarm fatigue} among caregivers. This in turn results in desensitization to alarms, noise disturbances and the possibility of missing a true life-threatening event lost among multiple alarms, a condition known as the \emph{cry-wolf} effect \cite{Sendelbach,Clifford_16}. The false alarms can also result in care disruption, sleep deprivation, patient anxiety, inferior sleep structure, and depressed immune systems \cite{imhoff2006alarm}. 
While the majority of current studies in this area focus on determining the optimal level of sensitivity for sensors, designing more accurate monitoring devices or more sophisticated data mining, and signal processing techniques to enhance the accuracy of false alarm detection using extracted information from individual monitoring devices, they often neglect the fact that most of the alarms triggered by individual sensors are considered false. This could be due to several factors including sensor detachment or motion artifacts. Therefore, extracting the correlation of information across different collected signals can play a significant role in identifying the false alarms\cite{CombineSig_Li2008,sadr2015reducing}.

One potential challenge of such correlation extraction among multiple collected signals is enhancing the computational complexity and the processing time of false alarm detection process as well as increasing the chance of over-fitting the trained model. Feature selection techniques can contribute to improving the prediction accuracy and reliability of such methods by removing irrelevant or redundant attributes across the big datasets. However, these methods usually evaluate individual contribution of the features and overlook their group impact when clustered together. Therefore, conventional feature selection techniques often discard the features that are highly correlated to the currently selected attributes, while these removed features can play a critical role in enhancing the accuracy of a model when grouped with other features.

The concept of coalition game theory has been recently applied to the feature selection problem as a means to capture the effect of grouping the features \cite{afghah2015game,Game_EMBC}. In these techniques, the impact of each feature is measured by calculating its Shapley value which is the average marginal contribution of each feature in enhancing the classification accuracy when it joins a coalition of selected features. However, the intensive computations involved in Shapley measurements make these methods impractical in predictive modeling applications with a large number of features. The estimation methods currently proposed to reduce this computational complexity, instead of calculating the Shapley value using all possible coalitions, only select a subset of these coalitions in a random manner. This approximation often compromises the performance of these techniques in applications where a high level of accuracy and reliability is expected. In this paper, we propose a genetic-algorithm-based method to estimate the Shapley value with a lower computational complexity in comparison to other Shapley estimation methods such as Monte-Carlo-based algorithms. In the proposed method, the most impactful coalitions of features are identified in a revolutionary process and are used to estimate the average impact of all coalitions. Such effective coalition sampling reduces the computational complexity of Shapley estimation by not calculating the impact of a large number of possible coalitions. Furthermore, in the previously reported game-theoretic based feature selection techniques, the contribution of each feature is measured based on its impact on enhancing the accuracy \cite{afghah2015game,Game_EMBC}. However, in false alarm detection and many other medical diagnosis applications, capturing the true positives is imperative. Therefore, enhancing the sensitivity is a more crucial factor to measure the performance of a predictive model. In this paper, we proposed a new metric to define the Shapley value of features that captures both sensitivity and specificity of a predictive model.

%The rest of this paper is organized as follow: in Section \ref{sec:data}, the dataset studied in this paper and the signal processing and feature extraction methods are described. Section \ref{sec:game} presents an introduction to coalition games and their prior applications in feature selection. In Section \ref{sec:GA_game}, the proposed genetic-algorithm based game theoretic feature selection method is described.

\section{Dataset Description} \label{sec:data}
In this study, we use the publicly available alarm dataset for ICUs by "PhysioNet computing in cardiology challenge 2015" that focuses on five life threatening arrhythmias including asystole, extreme bradycardia, extreme tachycardia, ventricular tachycardia, and ventricular fibrillation \cite{PhysioNet,Clifford_Data}. One objective of the proposed model is to reduce the rate of false alarms by considering the correlation among signals collected from different monitoring devices, therefore we considered 220 patients out of the entire training dataset with total of 750 patients for which three main signals of electrocardiogram II(ECG II), arterial blood pressure (ABP), and photoplethysmogram (PPG or PLETH) were available. The signals were re-sampled to 12 bit and 250 Hz and filtered by a Finite Impulse Response (FIR) bandpass [0.05 to 40 Hz] and mains notch filters for denoising. The alarms were labeled with a team of expert to either 'true' or 'false'. Among 220 reported alarms, 50 of those were true and the rest were false.

Motivated by the noticeable performance of discrete wavelet transform (DWT) in extracting informative time-frequency components of the physiological signals \cite{Saritha,Prochazka}, we applied this method to the three input signals of ECG II, ABP and PLETH. Six level decomposition using db8 for ECGII and db4 for ABP and PLET signals is utilized. Therefore, the three 1-D signals of each patient is converted into 18 vectors of wavelet coefficients. Since such transform generates a large number of wavelet coefficients that in turn can result in over-fitting of the trained model, we extract 20 statistical and information theoretical-based features of each wavelet vector coefficients. Some example features include mean, mode, median, range, variance, kurtosis, skewness, harmonic mean, interquartile range, Shannon entropy and log entropy. Moreover, in order to employ the Heart Rate Variation (HRV) information of the ECG II signals, a multi-resolution Wavelet technique is used to detect R-peaks of the signal \cite{banerjee2012delineation,jchen}. Afterward, the inverse R-R intervals which is so-called HRV signal is calculated and 20 statistical and information theoretical-based features of this HRV signal are extracted.
%\textcolor{red}{add a table with list of extracted features} 

 These 20 features are listed in Table \ref{tab:features}.

 \begin{table}[h]
 \centering
 \caption{Statistical and Information-theoretic features of wavelet vectors.}
 \label{tab:features}
 %\vskip 0.15in
 %\begin{center}
 %\begin{small}
 %\begin{sc}
 \begin{tabular}{|c|c|c|c|c|c|}
 %\hline
 %\multicolumn{2}{c}{Statistical} & \multicolumn{2}{c}{Inf. Theoretic}\\
 \hline
 No & Feature & No & Feature & No & Feature\\
 \hline
 1& mean & 8& std ($\sigma$) & 15 & Interquartile \\
 2& mode & 9& $\mu_3$ & & Range\\
 3& median & 10& $\mu_4$ & 16 & Shannon Ent. \\
 4& max & 11& coef. of var & 17 & Log Ent. \\
 5& min & 12& kurtosis & 18 & $n_T(max\{X_i\}/2)$ \\
 6& range & 13 & skewness & 19 & $n_T(max\{X_i\}/3)$ \\
 7& variance & 14 & H mean & 20 & $n_T(max\{X_i\}/4)$ \\
 \hline
 \end{tabular}
 %\end{sc}
 %\end{small}
 %\end{center}
 %\vskip -0.1in
 \end{table}
 
After extracting 380 statistical and information theoretical-based features of the wavelet coefficients and HRV signal, the feature sets of all the subjects are normalized. Considering the limited number of subjects compared to the number of features, we used a repeated k-fold method to evaluate the performance of the proposed feature selection model. In this experiment, we set $k = 5$ and repeated k-fold for 2 times by a random sampling manner, where created 10 copy of the database, each contains 175 observations in the training set and 45 observations in the test set. 

\section{Introduction to Coalition Games} \label{sec:game}
Cooperative (coalition) games refer to a class of game-theoretical models, where a cooperative behavior is enforced to the players in a way that the players prefer to form coalitions to obtain a higher payoff \cite{osborne2004introduction,Korenda_CISS}. Let us consider a finite non-empty set of players $\mathcal{I} = \{1,2,\ldots,n_f\}$, in which $n_f$ is the number of players and each player can participate in different sub-coalitions of $I$. The empty coalition is denoted by $\emptyset$ while the grand coalition, i.~ e.~ $\mathcal{I}$, is the coalition of all players. Also, the power set $P(\mathcal{I})$ is the family of all sub-coalitions of the grand coalition.

A cooperative game for the player set $\mathcal{I}$ is defined by a characteristic function $\nu: P(\mathcal{I})\rightarrow \mathcal{R}^+ \cup \{0\}$ with $\nu (\emptyset) = 0$, where $\nu(\mathcal{T} \in P(\mathcal{I}))$ represents the value of coalition $\mathcal{T}$. We use the notation $G(\mathcal{I} ,\nu)$ to represent all cooperative games on players in $\mathcal{I}$ with characteristic function $\nu$.
 A cooperative game $\nu$ is convex if for all $\mathcal{T}, \mathcal{S} \in P(\mathcal{I})$ we have $\nu(\mathcal{S} \cup \mathcal{T}) + \nu(\mathcal{S} \cap \mathcal{T}) \geq \nu(\mathcal{S})+\nu(\mathcal{T})$. The convex game $\nu$ is called super-additive if for all disjoint $\mathcal{S},\mathcal{T} \in P(\mathcal{I})$ we have $\nu(\mathcal{S} \cup \mathcal{I}) \geq \nu(\mathcal{S}) + \nu(\mathcal{T})$ \cite{osborne2004introduction}.
The \textit{marginal contribution} of player $i$ when it joins coalition $\mathcal{T} \subseteq \mathcal{I} \backslash \{i\}$ is defined as:
\begin{equation}\label{Def:MarginalDistr}
  \nabla_i(\mathcal{T} ) = \nu(\mathcal{T} \cup \{i\}) - \nu(\mathcal{T}).
\end{equation}

\emph{Shapley value} is a well-known solution concept for $\nu \in G(\mathcal{I} ,\nu)$ which measures the  marginal contribution of each player $i$ over all coalitions $\mathcal{T} \subseteq \mathcal{I} \backslash \{i\}$. Shapley function {$\phi: G(\mathcal{I},\nu) \rightarrow (\mathcal{R^+}\cup\{0\})^{P(\mathcal{I})}$, also called Shapley value on $G(I,\nu)$, needs to satisfy four axioms of coalition efficiency, dummy players, symmetry, and game additivity \cite{afghah2015game}. It has been proven that the following function $\phi:G(\mathcal{I},\nu) \rightarrow (R^+ \cup \{0\})^{P(\mathcal{I})}$ satisfies these aforementioned axioms:

\begin{equation}\label{Equ:ShapleyAccurate}
  \phi_i(\nu) = \sum\limits_{\mathcal{T} \subseteq \mathcal{I} \backslash\{i\}}\frac{|\mathcal{T}| (n_f - |\mathcal{T}| - 1)!}{n_f!}(\nu(\mathcal{T} \cup \{i\}) - \nu(\mathcal{T}))
\end{equation}

Coalition games have been recently applied to feature selection applications, where the features are considered as the players of the game\cite{Cohen_journal,afghah2015shapley,Game_BHI}. In these works, a coalition represents a group of features used for classification, where Shapley value of each feature measures the contribution of this feature in classification accuracy. Therefore, we can use Shapley value of each feature as its membership grade in the best coalition to identify the most salient features in the dataset. However, the considerable drawback of these methods is the associated computational complexity, because computing the Shapley value for each feature requires calculating the marginal contributions of that feature over all possible coalitions of any size. Therefore, these Shapley value-based methods either involve an intractable computational complexity for a large number of features or result in a degraded performance where a sub-group of all coalitions are randomly selected for Shapley calculation. In the next section, we propose a genetic-algorithm based method to distinguish an optimal set of coalitions to be utilized in estimating the features' Shapley values with low computational complexity and high accuracy.

\section{Proposed GA-based Monte-Carlo Method for Shapley Values Calculation} \label{sec:GA_game}
Noting the definition of Shapley value, the mathematical formulation of Shapely value of the $i$'th player presented in (\ref{Equ:ShapleyAccurate}) can be rewritten as:
\begin{equation}\label{Equ:Shepely}
  \phi_i(\nu) = \frac{1}{n_f} \sum_{t = 0}^{n_f-1} \underbrace{\frac{1}{\binom{n_f-1}{t}}
    \sum_{|\mathcal{T}| = t, i\not\in \mathcal{T} } [\nu(\mathcal{T}\bigcup\{i\}) -\nu(\mathcal{T})] }_{E(X_i^t)}
\end{equation}
where $E(X_i^t)$ is the average marginal contribution of player $i$ over all coalitions with size $t$ not including $i$ itself. This factor measures the effect of feature $i$ in classification accuracy when grouped with other features in different coalitions. The term average leads us to reducing computational complexity by operating Monte Carlo simulations over the $\binom{n_f-1}{t}$ possible coalitions of size $t$. Since there is no considerable correlation among the features in large-size coalitions; therefore we limit the calculation of marginal contribution of feature $i$ to the coalitions with size less than a specific threshold, i.~e.~ $n_f^{max} -1$. This in turn reduces the computational complexity of Shapley value calculation. Hence, the approximated shapely value of $i$'th feature can be written as:
\begin{equation}\label{Equ:ShapelyApprx}
  \hat{\phi}_i(\nu) = \frac{1}{n_f^{max}} \sum_{t = 0}^{n_f^{max}-1} E(X_i^t).
\end{equation}

In the following, we describe our proposed method to identify a subset of coalitions that provide higher marginal information in calculating Shapley value of user $i$.

\subsection{Proposed Genetic-algorithm based Shapley value calculation}
In order to estimate the Shapley value of each feature, we propose a genetic-algorithm (GA) method to generate the most effective subset of coalition sample sets. Such GA-based method involves defining proper chromosomes, fitness of each chromosome, and a revolutionary process of generating new generations. Moreover, parent selection, crossover and mutation are essential operations for a revolutionary process. The steps of the proposed GA are described in details as follows:

\paragraph{Chromosomes} The Shapley value estimation formula required an average on the marginal coalition values of the $i$'th feature over the coalitions of size $t$. Hence, we define each chromosome as a binary vector of length $n_f - 1$ which has exactly $t$ ones. By this, each chromosomes is mapped to a coalition with cardinality $t$.

\paragraph{Fitness Function}
While the majority of the current game-theoretic based feature selection methods only focus on enhancing the accuracy of classification in different supervised learning applications, one key contribution of our proposed feature selection method is to target elevating the Receiver-Operating Characteristic (ROC) curve as a measurement criterion for marginal contribution based on the rate history of the alarms. That enables us to not only increase the sensitivity of the classification but also enhance its specificity that is a particular interest to the false alarm reduction application.

To achieve this goal, the value of a coalition, i.e. $\nu(\mathcal{T})$, is proposed as a linear combination of specificity and sensitivity rates as defined in follow:
\begin{equation}\label{Equ:Coalval}
\nu(\mathcal{T}) = \frac{(1 - FNR_{\mathcal{T}}) + \mu (1 - FPR_{\mathcal{T}})}{1 + \mu},
\end{equation}
where $FNR_{\mathcal{T}}$ and $FPR_{\mathcal{T}}$ are the false negative and false positive rates obtained from the classifier, respectively and $\mu$ is a constant design parameter based on the history of the alarms. This model is appropriate for the imbalance data such as the available data for alarm dataset for ICUs.
% We make equal class variance assumption and create the Confusion Matrix (CM) of the test data using the Linear Discriminant Analysis (LDA). The CM from LDA test for coalition $\mathcal{T}$ can be obtained as:
% \begin{equation}\label{Equ:ConfMat}
%   CM = \begin{array}{|c|c|c}
%          \hline
%           TP & FN &P \\
%          \hline
%           FP & TN&N \\
%          \hline
%            P^* & N^*&
%        \end{array}
% \end{equation}
% where $P$ and $N$ are number of actual positive and actual negative classes in the test data, and $P^*$ and $N^*$ are number of predicted positive and predicted negative labels of the test data. $TP$, $FN$, $FP$, and $TN$ denote the number of positive predicted positive, positive predicted negative, negative predicted positive, and negative predicted negative, respectively. By using the definition of the confusion matrix in (\ref{Equ:ConfMat}), the coalition value of $\mathcal{T}$ can be rewritten as:
% \begin{equation}\label{Equ:Coalval}
%   \nu(\mathcal{T}) = \frac{TP}{P} + \mu \frac{TN}{N}.
% \end{equation}

Now, we define the fitness function of the proposed genetic algorithm, for a given feature $i$, the chromosome corresponds to the coalition $\mathcal{T}$ which does not include $i$, and a given coalition value as
\begin{equation}\label{Equ:fitness}
  f_i(\nu)(chr(\mathcal{T})) =  \nu(\mathcal{T}\cup \{i\} ) -  \nu(\mathcal{T}).
\end{equation}
In other word, the fitness of each chromosome for a given feature, is defined as the marginal value of the feature over the corresponding coalition.

%In follow, the method of population selection is described.

\paragraph{Parent Selection} For each feature, we randomly generate $n_p$ chromosomes, so called population, of the length $n_f - 1$ which each contains exactly $t$ ones. After calculating the population finesses, two chromosomes are being selected based on a random selection mechanism so called roulette mechanism. In the roulette mechanism, after normalizing the fitness set of population, a chromosome is selected with the probability proportional to the normalized fitness of the chromosomes in the population.

\paragraph{Crossover} In the crossover operation, two parents chromosomes are combined to generate two offsprings chromosomes such that those inherent path from both parents. In our proposed chromosome type, the crossover operation is done by finding non-unique same size chops of the parents chromosomes that locate in the same location, have the same number of ones, and have a length greater than one; then we randomly select one of those chops and exchange the chops between two parents chromosomes. However, it is possible that such chops do not exist in the parents chromosomes. In that case, each parents chromosomes is updated via a hermaphrodite cross over operation in which a randomly selected chop of chromosome is chosen and after reversion, fit back to its location in the chromosome.

\paragraph{Mutation} In most revolutionary techniques, some randomness is required to obtain the diversity in the field search. We consider mutation of one bit 0 and one bit 1 in each offsprings' chromosomes. After mutation we add the generated offsprings' chromosomes to the population, calculate their corresponding fitness, and update population by removing two chromosomes with lowest fitness from it. However, we keep those chromosomes as a valid sample set for estimating the Shapley value of the $i$'th feature.

 In follow, we discuss the relation between statistical properties of the samples obtained from GA and statistical properties of all possible feature coalitions with size $t$.

\subsection{Mean Adjustment of Samples}
The proposed GA algorithm for generating coalition samples tends to select the chromosomes with highest marginal contribution for $i$'th feature. The marginal contribution of the selected coalitions for feature $i$ can be modeled as random variable $Y_i^t$ which is the maximum among M marginal contribution samples of all size-$t$ coalitions, i.~e.~ $\{X_{i,m}^t\}_{m = 1}^{M}$. This relation can be written as $Y_i^t = \max\{X_{i,m}^t\}_{m = 1}^{M}$.
The samples $X_{i,m}^t$ are independent, so the Cumulative Density Function (CDF) of random variable $Y_i^t$, when $M >> 1$, can be written as $F_{Y_i^t}(y) = \exp(-\exp(-\frac{y-u}{\alpha})), \;\;\; -\infty <y<\infty, \alpha>0$ \cite{jowitt1979extreme}, and the distribution of $Y_i^t$ is called Extreme Type 1 (EX1) distribution. The parameters $\alpha$ and $u$ of the EX1 distribution are the root square variance and mean of the distribution $X_i^t$. The expected value and variance of EX1 can be estimated as $E\{Y_i^t\} = u + {0.5722}\alpha$ and $Var\{Y_i^t\} = {1.645}{\alpha^2}$.

Assuming $n_G$ is the number of generated samples from GA such that $n_G << \binom{n_f-1}{t}$, then $M = \lfloor \frac{\binom{n_f-1}{t}}{n_G}\rfloor$ is large enough and we can use the above mentioned approximation.}, which obtained from GA, by EX1 distribution. Therefore, by extracting the statistical information of the samples $Y_i^t$, the mean (and variance) of all marginal information of size $t$ coalitions for feature $i$ will be estimated. Finally, the $n_f$ features with highest Shapley values are selected for the classification purposes. 
In the next section, we analyze the complexity of the proposed feature selection algorithm.

\section{Complexity Analysis}
The Shapley value based feature selection methods involve an exponential computational complexity that make them being classified as NP-hard problem. Hence, feature selection methods based on calculating Shapley values of the features are computationally intractable when the number of features is very large. However, one may reduce the complexity order of this process by limiting the size of feature coalitions which are considered for Shapley value calculation \cite{afghah2015shapley}. In that scenario, the complexity of the algorithm is reduced to $O(n_f^{n_f^{max}})$, however the performance is also degraded. One considerable advantage of our proposed method compared to previously reported game-theoretic based feature selection techniques is a lower computational complexity in estimating Shapley value by employing a GA-based Monte-Carlo method. This method reduces the complexity order of the estimation to $O(n_f \times n_f^{max} \times n_G)$.

\section{Numerical Results}
In this section, we present the numerical results to evaluate the performance of the proposed GA-based method in identifying the salient features. We used the publicly available alarm dataset for ICUs from PhysioNet challenge 2015 and extracted 380 features for each patient as described in Section \ref{sec:data}. 
In the proposed method, we measured the impact of each feature over coalitions with size of less than 20 by getting the average of marginal contributions over 100 coalitions of each size that are selected by the proposed genetic algorithm. The Shapley value of each feature is then estimated by finding the average of the obtained marginal contributions for all coalition sizes. This process is repeated for three different values of $\mu = 0.5, 1.0$ and $3.5$ in (\ref{Equ:Coalval}). The 20 features with highest Shapley values for each $\mu$ are selected for classification purposes. In Table \ref{tb:FeatureSelectionPerformance}, the performance of the proposed feature selection method is compared with several popular feature selection methods including $\chi_2$, Tree-based method in which forest of trees are used for calculating feature values \cite{pedregosa2011scikit}, Mutual Gain Information, Relief, and three types of Wrapper feature selection approaches. The output of each feature selection method is then evaluated using different classifiers including decision trees, discriminant analysis, logistic regression, Support Vector Machine (SVM), Nearest Neighbors, and ensemble classifiers. However among different classification methods, the RUSBoosted Trees Ensemble method is reported since higher sensitivity values for feature selection methods are achieved.

\begin{table*}
	 % \small
	\centering
	\caption{Comparison of classification performance for different feature selection methods with best classifiers in terms of accuracy and/or AUC}
	\resizebox{0.8\linewidth}{!}{  %fit to windows command 
		\begin{tabular}{|c|c|c|c|c|} \hline
			Feature Selection  & Accuracy &  AUC & Sensitivity & Specificity \\ \hline
			 
			Shapley $\mu = 3.5$  &0.77 & 0.81 & {\textbf{0.73}} & 0.75 \\ \hline
            Shapley $\mu = 1.0$  & 0.75 & 0.80 & 0.72 & 0.75\\ \hline
              Shapley $\mu = 0.5$ &0.76 & 0.80 & 0.70 & 0.77\\ \hline
			$\chi_2$    & 0.71& 0.77 & 0.71 & 0.72\\ \hline
			Tree Based   & 0.75& 0.79 & 0.66 &0.78 \\ \hline
    Mutual Gain Information  & 0.76& 0.84 & {\textbf{0.73}} & 0.75 \\ \hline
  Relief  & 0.81 & 0.77 & 0.60 & 0.87 \\ \hline  
Wrapper: LASSO  & 0.76 & 0.82 & 0.66& 0.79  \\ \hline  
Wrapper: Ridge Regression  & 0.73 & 0.77 & 0.62& 0.76   \\ \hline
Wrapper: Logit Regression  & 0.75 & 0.76 & 0.62 & 0.78   \\ \hline
		\end{tabular}%
	}
	\label{tb:FeatureSelectionPerformance}%
\end{table*}

 This result also shows a balance between the sensitivity and specificity of the proposed model that can be obtained by tuning the $\mu$ value. As it can be seen from this table, the obtained sensitivity from the proposed method has highest value among all other feature selection methods. 

\begin{figure}
%\begin{center}
%\centering
 \centerline{
 \resizebox{!}{6.7cm}{\includegraphics{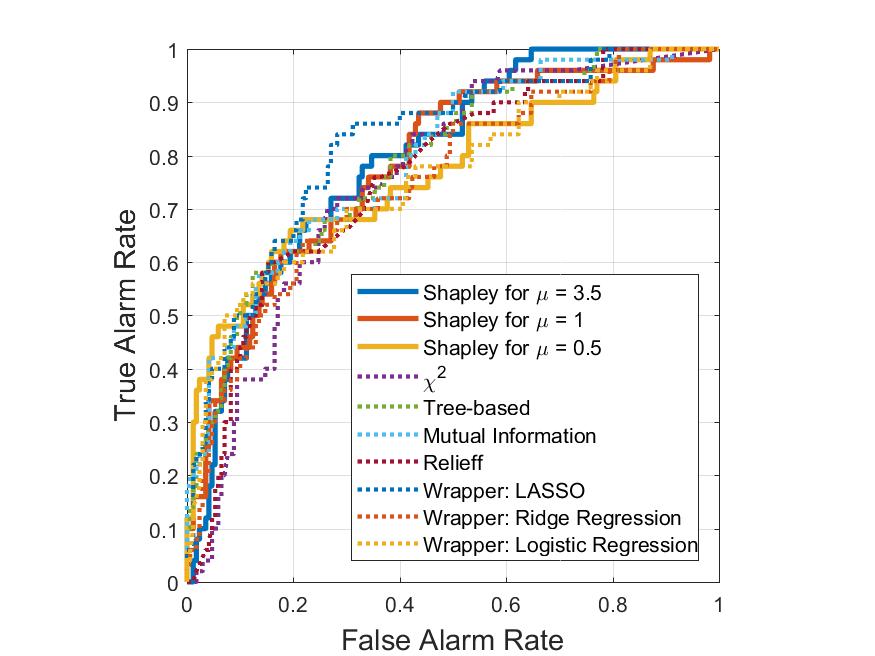}}
 }
\caption{The ROC of different feature selection methods with their best classification in terms of AUC.}
\label{fig:ROCCurves}
%\end{center}
\end{figure}

Figure \ref{fig:ROCCurves} compares the ROC curves of the proposed feature selection with other feature selection approaches. As it is shown in this figure, the ROC of the Shapley method with $\mu = 3.5$ has highest values after false alarm 0.6, and the curve is above most of the other ROC's for false alarms less than 0.6.

Another aspect of our work is employing different biomedical signals and different signal processing types (Wavelet and HRV for ECG II signals) for the purpose of increasing the classification performance. Table \ref{tb:FreqAnalyseBasedOnSignal} shows the frequency analysis of the number of features which are selected during different feature selection approaches from different biomedical signals or signal processing type. One of the interesting results from this is that the proposed algorithm which selects features with high marginal contributions, selects more features from Wavelet features than the HRV features. This table also shows that employing different source of biomedical signal is useful. It can be also seen that most of the selected features in the proposed algorithm are from ECG II Wavelet features and the PLETH Wavelet features. It can be justified with the sense that if the number of signal sources increased, the chance of adding more correlated features is also increased.

\begin{table}
	  %\small
      \normalsize
      \centering
	\caption{Frequency analysis of selected features based on signal types and signal processing types for different feature selection techniques.}
	\resizebox{0.9\linewidth}{!}{  %fit to windows command 
		\begin{tabular}{|c|c|c|c|c|c|} \hline
			Feature Selection &  Total & $\begin{array}{c}
			\mbox{ECG II} \\
            \mbox{Wavelet}
			\end{array}$ &
            $\begin{array}{c}
			\mbox{PLETH} \\
            \mbox{Wavelet}
			\end{array}$ &
            $\begin{array}{c}
			\mbox{ABP} \\
            \mbox{Wavelet}
			\end{array}$ &
            $\begin{array}{c}
			\mbox{ECG II} \\
            \mbox{HRV}
			\end{array}$ 
            
            \\ \hline 
            Shapley $\mu = 0.5$
             &20&0&20&0&0
			 \\ \hline
            Shapley $\mu = 1.0$
             &20&5&15&0&0
			 \\ \hline
            Shapley $\mu = 3.5$
             &20&6&14&0&0
			 \\ \hline
            $\chi_2$ &20&0&14&6&0
			 \\ \hline
            Tree Based &139&47&53&33&6
			 \\ \hline
             Mutual Gain Information &20&17&0&0&0 
			 \\ \hline
            Relief &20&2&16&2&0 
			 \\ \hline
            Wrapper: LASSO &25&9&11&4&1
			 \\ \hline
            Wrapper: Ridge Regression &136&44&57&26&9 
			 \\ \hline
            Wrapper: Logit Regression &148&50&55&32&11
			 \\ \hline
		\end{tabular}%
	}
	\label{tb:FreqAnalyseBasedOnSignal}%
\end{table}

\section{Conclusion}
In this paper, a low-complexity feature selection method for false alarm reduction in ICUs is proposed, where the Shapley values of the features extracted from physiological signals are estimated through a GA-based algorithm. These Shapley values evaluate the impact of grouping of multiple features in enhancing sensitivity and specificity of the trained model. The numerical results show that the specificity of this propose method is comparable to other existing feature selection methods while it offers a higher sensitivity as desired in alarm detection application to assure capturing the true alarms. 

\bibliographystyle{IEEEtran}
\bibliography{IEEEabrv,reference}
\end{document}